\documentclass{emulateapj}

\shorttitle{Transient jets} \shortauthors{Pe'er \& Markoff}

\newcommand{\cm}{\rm{\, cm}}

\newcommand{\eV}{\rm{\, eV }}
\newcommand{\keV}{\rm{\, keV }}

\newcommand{\erg}{\rm{\, erg }}

\newcommand{\beq}{\begin{equation}}
\newcommand{\eeq}{\end{equation}}
\newcommand{\ba}{\begin{array}}
\newcommand{\ea}{\end{array}}

\newcommand{\D}{{\mathcal {D}}}

\def \etal{{\it et al.~}}

\begin{document}
\title{X-ray Emission from Transient Jet Model  in Black Hole Binaries} 

\author{Asaf Pe'er\altaffilmark{1}, Sera Markoff\altaffilmark{2}}

\altaffiltext{1}{Harvard-Smithsonian Center for Astrophysics, 60
  Garden Street, Cambridge, MA 02138, USA} 
\altaffiltext{2}{Astronomical Institute "Anton Pannekoek", University
  of Amsterdam, P.O. Box 94249, 1090 GE Amsterdam, The Netherlands} 

\begin{abstract}
  While the non-thermal radio through at least near-infrared emission in
  the hard state in X-ray binaries (XRBs) is known to originate in
  jets, the source of the non-thermal X-ray component is still
  uncertain.  We introduce a new model for this emission, which takes
  into account the transient nature of outflows, and show that it can
  explain the observed properties of the X-ray spectrum. Rapid
  radiative cooling of the electrons naturally accounts for the break
  often seen below around 10 keV, and for the canonical spectral slope
  $F_\nu \propto \nu^{-1/2}$ observed below the break. We derive the
  constraints set by the data for both synchrotron- and
  Compton-dominated models. We show that for the synchrotron-dominated
  case, the jet should be launched at radii comparable to the inner
  radius of the disk ($\sim$few 100 $r_s$ for the 2000 outburst of XTE
  J1118+480), with typical magnetic field $B \gtrsim 10^{6}$~G. We
  discuss the consequences of our results on the possible connection
  between the inflow and outflow in the hard state of XRBs.
\end{abstract}


\section{Introduction}
\label{sec:intro}

The flat/inverted radio spectrum seen in the hard state of black hole
X-ray binaries \citep[XRBs, e.g.,][]{Hynes00, Fender01, Fender06}, is
naturally accounted for by self-absorbed synchrotron emission from
jets \citep{BK79, HJ88, FB99}, analogous to the compact radio cores of
active galactic nuclei (AGN).  Although generally too small to
resolve, the compact jets have been imaged in the case of Cyg X-1
\citep{Stirling+01}, confirming the origin of the emission.  In
contrast, the origin of the hard, non-thermal X-ray component is more
uncertain.  On the one hand, it has long been known that
accretion-based models (inverse Compton emission from the inner parts
of the disk) can have a significant contribution to the hard X-ray
spectrum \citep[e.g.,][]{Titarchuk94, MZ95, Gierlinski+97, Esin97,
  Esin01, Poutanen98, Cadolle+06, YZXW07}.  On the other hand, it is
also possible that jets can contribute significantly to the
observational properties at these bands \citep{MiRo94, MFF01, MNCFF03,
  MNW05, BRP06, PBR06, Kaiser06, GBD06, Kylafis+08, Maitra+09}.
Indeed the emission from jets in other objects, such as AGN, can span
decades in spectral frequency, from the radio all the way to the X-
and $\gamma$-ray bands.

While both scenarios can provide equally good fits to the X-ray
spectrum \citep[see, e.g.,][]{Nowak+11}, the thermal corona inverse
Compton (TCIC) classes of models do not generally connect well to the
need for strong magnetic fields and outflows in the same region,
implied by the necessity of launching collimated jets.  Because of the
very high temperatures in the inner disk regions \citep[$\sim 1$
keV;][]{SS73}, the hard X-ray emitting plasma will be mostly ionized,
and thus be subject to electromagnetic forces and in potentially near
relativistic motion.  This motion together with the size of the
dominant emitting region implies a characteristic dynamical time that
can play an important role in the radiative properties of the system,
and leave its mark in the spectrum.

Additional clues come from the spectral behavior at the X-ray band.
In several outbursts, the X-ray spectrum shows a break between a few
to $\sim$~tens of keV, where the spectral index steepens from
$\alpha\sim0.5$ (where $F_\nu \propto \nu^{-\alpha}$) to
$\alpha\sim1$\footnote{Often the data at these energies are noisy,
  hence the errors in the values of $\alpha$ after the break are
  fairly large.}.  Such a break can be seen in numerous objects.  For
example, it is seen at 2 keV in the 2000 outburst of XTE~ J1118+480
\citep{Hynes00, Esin01}, at $\lesssim$~40 keV in the 2000 outburst of
XTE~ J1550-564 \citep{RCT03}, at $\sim$~70 keV in the 2002 outburst of
GX~339-4 \citep{Homan+05}, at $\approx$~40 keV in the 2005 outburst of
GRO~J1655-40 \citep{JKS08}, at $\sim 40 \keV$ in the 1991 outburst of
Cyg X-1 \citep{Gierlinski+97} and at $\sim 7$~keV in the 2002 outburst
of GRB 1758-258 \citep{Soria+11}, to name only few examples.  These
breaks are most pronounced in the 'hardest' state. Often, these breaks
disappear and the spectrum becomes softer with the increase of the
luminosity, and the transition to the high/soft state
\citep[e.g.,][]{Dunn+11}.

The existence of this class of spectral breaks is not easily explained
in the framework of current models.  Several ideas include Compton
reflection by cold matter, which is expected to harden the X-ray
spectrum above $\sim 10$~\keV \citep{LW88, HM93, MZ95}.  Another idea
is metal absorption by partially ionized gas \citep[e.g.][]{Esin01}.
Alternatively, the break may result from combination of synchrotron
spectrum at low energies, and synchrotron self-Compton at higher
energies \citep[e.g.,][]{MNW05,Homan+05}.

While the above mentioned suggestions may indeed play a role in
producing the break, we note that the low energy spectral index,
$\alpha \simeq 0.5$, may hold the key to understanding the
emission. Indeed, this spectral index is a natural outcome of
both synchrotron and inverse-Compton emission processes, provided that
the radiating particles cool on a time scale much shorter than the
dynamical time scale.  This fact together with the need to account for
plasma bulk motion thus motivates us to suggest an alternative model
for the hard X-ray spectrum in hard state XRBs.

As we show below, the observed X-ray spectral properties of many of
these objects can follow directly from the acceleration of electrons
at some radius $r$. While it is clear that particles are accelerated
to high energies in many astronomical objects, the details of this
process are still highly uncertain.  From a theoretical perspective,
it has long been argued that Fermi acceleration in shock waves
\citep{Bell78, BO78, BE87}, or possibly in magnetic reconnection
layers \citep[e.g.,][and references therein]{Kirk04} is a plausible
mechanism.  Many works study this phenomena in both the Newtonian
regime \citep[e.g.,][]{AB05}, and the ultra-relativistic regime
\citep[e.g.,][]{Spit08, SS11}. These works suggest that indeed
electrons are accelerated in shock waves into a power law distribution
above a characteristic Lorentz factor, $\gamma_m$. However, it is
found that in both regimes only about $\sim 1\%$ of the particles are
accelerated to a power law, the rest maintain a low energy Maxwellian
distribution. A similar result was found in studies of particle
acceleration in supernovae remnants \citep[SNRs;
see][]{Ellison+07}. The emerging theoretical picture is therefore that
particles are heated as they cross the shock wave, and their spectra
is composed of a relativistic Maxwellian distribution with a small
fraction accelerated to a power law distribution at higher energies
\citep[see also][]{Caprioli+10}.

On the other hand, phenomenological works of spectral fitting in
various astronomical objects \citep[e.g., astrophysical jets,
supernovae remnants - ][]{Lazendic04} and gamma-ray burst (GRB)
afterglows \citep{G98b, WG99} suggest that the power law tail of the
electron distribution dominates.  Thus, there is a discrepancy between
the phenomenological works and the theoretical predictions.

Regardless of the exact details, following the acceleration the
electrons radiate their energy via synchrotron emission and
inverse-Compton scattering of both the disk (thermal) UV/X-ray photons
and locally produced synchrotron photons (synchrotron self-Compton;
SSC).  As the electrons radiate their energy they cool.  Interestingly
enough, it was shown by \citet{PC09}, that the only requirement needed
in order to obtain the universal spectrum observed below the break,
$F_\nu \propto \nu^{-1/2}$, is that the cooling time scale is shorter
than the dynamical time scale (the particles are in the ``fast
cooling'' regime). Thus, this result is independent on the uncertain
details of the acceleration process.

The spectral shape above the break does depend on the details of the
acceleration process. Due to the uncertainty mentioned above, in the
calculations presented below we consider both scenarios, the
theoretically-motivated scenario in which most accelerated particles
have a Maxwellian distribution, and the phenomenologically-motivated
scenario, in which the spectral shape of most accelerated particles is
a power law.

Fast cooling of a power-law distribution of electrons leads to a
spectral index $F_\nu \propto \nu^{-p/2}$ above the spectral break,
where $p$ is the power law index of the accelerated electrons. Thus,
if most electrons are accelerated to a power law distribution, then
the observed spectral slopes at high energies, $\alpha \simeq 1$, are
therefore consistent with the nearly universal power law index $p =
2.2 \pm 0.2$ observed in many astronomical objects. As we claim here,
the derived conditions at the emission site, such as emission radius
and magnetic field strength are most naturally accounted for in
turbulent outflows, as would be expected for a region of jet
launching.

If, on the other hand, most accelerated electrons have a Maxwellian
distribution, than a power law spectrum above the break is not
expected. However, as we show below, the finite width of the
Maxwellian distribution makes it possible to fit the observed spectra
in this scenario as well. In this case, the existing data does not
provide information on the power law index of the most energetic
electrons.

We therefore present here a new jet model that can account for the
X-ray properties of many X-ray binaries. By solving the kinetic
equations that govern particle and photon populations
self-consistently, we are able to reproduce the main observational
properties of the X-ray spectra. In particular, we suggest a natural
explanation for the observed $\alpha=0.5$ to $\alpha\sim1$ spectral
break as resulting from rapid radiative cooling of the energetic
particles. Our numerical calculations are based on the model developed
by \citet{PW05} in the study of the prompt emission of GRBs, that was
further modified in the study of the radio properties of X-ray
binaries \citep{PC09}. Here, because we focus on the X-ray properties
of XRBs, various physical processes that were not required for radio
emission studies are now included, such as Compton scattering and pair
production and annihilation.  

This paper is organized as follows:  In \S\ref{sec:model} we present
the basic properties of our model. In particular, we focus on the
rapid cooling that leads to the universal $F_\nu \propto \nu^{-1/2}$
spectra below the spectral break. We compare our model to the data in
\S\ref{sec:results}, before summarizing and discussing our results in
view of the recent observations in \S\ref{sec:summary}.

\section{Model assumptions and basic properties}
\label{sec:model}

Similar to earlier jet models \citep[e.g.,][]{MFF01,MNW05, Maitra+09},
we assume that the initial gas which is transferred from a mass-losing
secondary forms a geometrically thin, optically thick cool disk
\citep{SS73}. Emission from the disk dominates the emission at the
extreme-ultraviolet and soft X-ray bands.  Thus, fitting the data at
these energy bands provides an estimate of the thermal luminosity
$L_{th}$ and the temperature $T_d$ of particles and photons in the
inner part of the disk, and is often seen as a proxy to the bolometric
luminosity of the system.  The inner disk radius is then given by
$r_{in} \simeq (L_{th}/2\pi \sigma_S)^{1/2} T_d^{-2}$, where
$\sigma_S$ is the Stefan-Boltzmann constant.

There is still some debate over whether the data from hard state XRBs
implies a truncated inner disk radius or not, with often conflicting
results from continuum fitting of the thermal accretion disk, and the
modeling of the potentially broadened iron fluorescence line at
6.4~keV.  Several groups claim an inner radius of typically $r_{in}
\sim 10^{2} - 10^3 \, r_s$, where $r_s =2G M_{BH}/c^2$ is the
Schwarzschild radius of a non-rotating black hole \citep{Esin01,
  Tomsick+09, DDT10}.  On the other hand, several recent works argue
that the data imply a lower inner radius, $\sim 10 - 10^{1.5} \, r_s$,
\citep{RMHM10}, or even against the need for a recessed disk to
explain the data \citep[e.g.,][]{RMF09, RM10}.

Regardless of the exact radius where it occurs, generally we do expect
a transition of properties in the inflow/outflow of the accreting
plasma.  Once the gas approaches the disk inner radius $r_{in}$, it is
expected to become geometrically thicker and optically thin, and also
radiatively inefficient.  Several variations of the original
advection-dominated accretion flow \citep[ADAF; see][for a recent
review]{NM08} scenario, more generally referred to as radiatively
inefficient accretion flows (RIAFs) now exist.  Because RIAFs are
likely necessary to support jet launching \citep[see, e.g.,][]{LOP99,
  Meier01}, we here assume such a transition is likely.  However,
since emission from RIAFs has been extensively studied in the past
\citep[e.g.,][]{Esin97, Esin01}, we currently focus on the jets.
While we assume here that the jets dominate the
emission\footnote{Similar assumptions were taken by, e.g.,
  \citet{YMF02, YC05}} , we note that a possible contribution from the
RIAF may alter our final results.

We thus assume that at some radius $r_j$, part of the plasma is
ejected outward from symmetric nozzles into a jet outflow. In the
calculations presented, $r_j$ is taken as an independent variable,
that is not necessarily equal to $r_{in}$.  During the ejection
process, the electrons are accelerated. Such an acceleration can
result from Fermi acceleration in shock waves \citep{EJR90, Spit08},
or perhaps in reconnection layers of strong magnetized outflows
\citep[e.g.,][]{DS02}. We parameterize the current uncertainties about
the details of the acceleration process by simply assuming that a
fraction $\xi_{pl}$ of the electrons are accelerated to a power law
distribution with power law index $p$ above a characteristic Lorentz
factor $\gamma_m$, likely set by the thermal temperature of the
innermost accretion flow. The remaining electrons (a fraction
$1-\xi_{pl}$) assume a relativistic Maxwellian distribution, with
temperature $\theta \equiv kT/m_e c^2 = \gamma_m/2$, ensuring smooth
connection with the power law at higher energies.

In order to connect the acceleration microphysics to the system
energetics, we further assume that a fraction $\epsilon_e$ of the jet
kinetic luminosity $L_k$ is dissipated away, carried by the energetic
electrons.  This assumption therefore results in the constraint
\citep[e.g.,][]{PW04}:
\beq 
\gamma_m = \left\{ \ba{ll} \epsilon_e
  \left({m_p \over m_e} \right) \ln^{-1} \left({\epsilon_{\max} \over
      \epsilon_{\min}}\right), & p=2,
  \nonumber \\
  \epsilon_e \left({m_p \over m_e} \right) {p-2 \over p-1}, & p\neq2.
  \ea \right.
\label{eq:gamma_m}
\eeq 
Here, $\epsilon_{\max}$ ($\epsilon_{\min}$) is the maximum (minimum)
energy of the accelerated electrons . The maximum energy is calculated
by equating the acceleration time to the cooling time (e.g., via
synchrotron emission), and the minimum energy, for power law index
$p=2$ is calculated iteratively, as $\epsilon_{\min} = \gamma_m m_e
c^2$. Following the calculation of $\gamma_{m}$, it is corrected
iteratively to account for the fraction $1-\xi_{pl}$ of the electrons
that assume a Maxwellian distribution, so that the energy density in
the combined (Maxwellian + power law) electron population exactly
equals a fraction $\epsilon_e$ of the energy density in the jet.

The particles propagate at bulk velocity $\beta_j c$ inside the jet,
during a (comoving) dynamical time $t_{dyn} \simeq \gamma_j
r_j/\beta_j c$, where $\gamma_j = (1-\beta_j^2)^{-1/2}$ is the Lorentz
factor associated with the bulk motion of the plasma. During this
time, a continuous injection (and acceleration) of particles at the
base of the jet is assumed.  We assume the existence of a steady
magnetic field of strength $B$ inside the jet. While in some works the
magnetic energy density is described as a fraction $\epsilon_B$ of the
jet energy (analogue to the definition of $\epsilon_e$), we note that
as the source of the magnetic field may be attached to the inner parts
of the disk, it is possible that $\epsilon_B \geq 1$. Therefore, here
we consider $B$ to be a free parameter.
As a result, once introduced into the jet, the particles
lose their energy via radiatively producing energetic photons. This
energy loss results from synchrotron emission and by inverse Compton
scattering of both the thermal disk photons as well as the synchrotron
emitted photons (SSC).

The existence of a characteristic physical scale $r_j$ (corresponding
to characteristic time $t_{dyn}$) implies that the physical quantities
(e.g., the magnetic field $B$ or the energy density in the photon
field) do not vary significantly over this scale. In particular, we
neglect adiabatic energy losses over this scale, as its contribution
to the electrons cooling is (by definition) much weaker than the
radiative cooling \citep[see further discussion in][]{PC09}.

The radiative cooling time of electrons at energy $E = \gamma m_e c^2$
is given by  (in the Thompson regime):
\beq
t_{cool} \simeq {E \over P} = {\gamma m_e c^2 \over (4/3) c \sigma_T
  \gamma^2 (u_{th} + u_B)},
\label{eq:t_cool}
\eeq 
where $u_{th} = L_{th}/\pi r_j^2 \beta_j c \gamma_j^2$ and $u_B =
B^2/8 \pi$ are the (comoving) energy densities of the thermal photons
(assuming cylindrical geometry) and the magnetic field, respectively.
We denote by $\gamma_c$ the Lorentz factor of the electrons for which
the cooling time is comparable to the dynamical time, $\gamma_c = (m_e
c^2)/(4/3) c \sigma_T (u_{th} + u_B) t_{dyn}$. For $\gamma_c \ll
\gamma_m$, the electron energy distribution in the range $\gamma_c \ll
\gamma \ll \gamma_m$ reaches a steady state, which is calculated by
solving the rate equation, ${dn_{el}(\gamma,t) / dt} = (\partial
/ \partial \gamma) \left[ n_{el}(\gamma) {\partial \gamma / \partial
    t} \right] = 0.$ Since the power emitted by both synchrotron
radiation and Compton scattering is $P_{syn, IC} \propto \partial
\gamma /\partial t \propto \gamma^2$, the steady state electron
distribution at the range $\gamma_c \ll \gamma \ll \gamma_m$ is
$n_{el} (\gamma) \propto \gamma^{-2}$. Above $\gamma_m$, the steady
particle distribution (assuming an initial power law) is
$n_{el}(\gamma) \propto \gamma^{-(p-1)}$. Therefore, the obtained
spectral shape from either synchrotron- or inverse Compton-dominated
cooling is
\beq
F_\nu \propto \left\{
\ba{ll}
\nu^{1/2} & \nu \ll \nu_m, \nonumber \\
\nu^{-p/2} & \nu \gg \nu_m, 
\ea
\right.
\label{eq:F_nu}
\eeq
where $\hbar \nu_m = \varepsilon_{\min}$ is the typical energy of photons
emitted by electrons at $\gamma_m$.  

Equation \ref{eq:F_nu} is valid if most electrons are accelerated to a
power law. If $\xi_{pl} \ll 1$, then above the break at $\nu_m$ a
smooth variation in the spectrum is expected, resulting from the
finite width of the Maxwellian distribution (see below).

The solution obtained in equation \ref{eq:F_nu} is valid only as long
as $\gamma_c-1 \gtrsim 1$ (and $\gamma_c \ll \gamma_m$). If the
cooling is extremely rapid, the electrons become sub-relativistic on a
time scale much shorter than the dynamical time. Under these
conditions, the sub-relativistic (cold) electrons interact with the
low energy (disk) photons via multiple Compton scatterings, altering
their original distribution \citep[for discussion on this effect in
the context of GRB prompt emission, see][]{PMR05, PMR06}. Thus, a
sharp cutoff of the disk photons can be used to constrain the optical
depth, hence discriminate between jet and accretion-based models. See
further discussion in \S\ref{sec:summary} below.

\section{Constraints set by observations}
\label{sec:results}

The broad band observations of the XRB hard state contain a wealth of
data that can be used to provide strong constraints on the free model
parameters. The rapid cooling of the electrons implies that nearly
100\% of the energy given to the electrons is converted to emission,
mainly at the X-ray band. Thus, a constraint on the jet kinetic
luminosity as well as the fraction of energy used to accelerate the
electrons is $L_{NT} \simeq L_k \epsilon_e$, where $L_{NT}$ is the
luminosity in energetic photons (with energies above that of the disk
photons). The break energy at few keV corresponds to
$\varepsilon_{\min}$, and, for $\xi_{pl} \lesssim 1$, the spectral
slope above the break provides a direct indication of the power law
index $p$ of the accelerated particles via eq. \ref{eq:F_nu}.

A large optical depth to scattering by the electrons and the produced
pairs will result in a modification in the emission from the inner
parts of the disk. Thus, the sharp cutoff in the disk emission implies
that the optical depth to scattering cannot be much larger than unity,
$\tau_{\gamma e} \lesssim 1$. An order of magnitude estimate of the
optical depth is obtained by assuming cylindrical jet geometry with
scale $r_j$. The comoving number density of injected electrons is
$n'_{el} = L_k/\pi r_j^2 m_p c^2 \gamma_j^2$, resulting in optical
depth $ \tau_{\gamma e} \simeq\gamma_j r_j n'_{el} \sigma_T = {L_{NT}
  \sigma_T / \pi r_j \epsilon_e m_p c^3 \gamma_j \beta}$.

The optical depth is modified by the existence of pairs.  An estimate
of the optical depth in the vicinity of pairs can be done as follows
(a detailed description will be presented in another work).  For short
dynamical times, pair annihilation is insignificant, and the number
density of pairs is $n_{\pm} \simeq n_{\gamma(\varepsilon>m_e c^2)}
n_{\gamma(\varepsilon<m_e c^2)} c \sigma_T t_{dyn}/ 4$.  For a flat
spectral energy slope ($\nu F_\nu \propto \nu^0$) over the energy
range $\varepsilon_{\min}$--$\varepsilon_{\max}$ (as is expected for
electrons power law index $p$ not much different than 2), the number
density of photons at the different energy bands are
$n_{\gamma(\varepsilon>m_e c^2)} \simeq u_{NT}/m_e c^2
\log(\varepsilon_{\max}/\varepsilon_{\min})$, and
$n_{\gamma(\varepsilon<m_e c^2)} \simeq n_{\gamma(\varepsilon>m_e
  c^2)} \times (m_e c^2/\varepsilon_{\min})$, where $u_{NT} =
L_{NT}/\pi r_j^2 c \beta \gamma_j^2$ is the energy density in the
non-thermal component. Combined together, the optical depth for
scattering by the pairs is
\beq
\ba{lcl} 
\tau_{\gamma,e^\pm} & \approx & 
\left({L_{NT} \sigma_T \over 2 \pi r_j \gamma_j \beta_j^2 m_e
    c^3}\right)^2 \beta_j {m_e c^2 \over \varepsilon_{\min}
  \log\left(\varepsilon_{\max} / \varepsilon_{\min}\right)^2}
\nonumber \\
& \simeq &
{1 \over \beta_j^3 \gamma_j^2} \left({L_{NT} \over 5\times 10^{-3}
    L_E}\right)^2 \left({r_j \over 10 r_s}\right)^{-2}.
\ea
\label{eq:tau}
\eeq 
In estimating the optical depth in equation \ref{eq:tau}, we used
parameters that characterize XTE J1118+480, such as distance $d =
1.72$~kpc \citep{Gelino+06}. We estimate $M_{BH} \simeq 7
M_\odot$\footnote{While \citet{McClintock+01} estimated $M_{BH} = 6.1
  M_\odot$, \citet{Gelino+06} obtained $M_{BH} = 8.5M_\odot$.},
resulting in Eddington luminosity $L_E = 8.75 \times 10^{38} {\rm \,
  erg \, s^{-1}}$ . Since $\log(\varepsilon_{\max}/\varepsilon_{\min})
\approx 10$, for $\varepsilon_{\min} \simeq {\rm few} \keV$,
$m_ec^2/\varepsilon_{\min} \log(\varepsilon_{\max}/\varepsilon_{\min})
\approx few$.  Thus, the requirement $\tau_{\gamma,e^\pm} \lesssim 1$
implies, for mildly relativistic jets (say, $\beta_j \approx 0.4$),
$r_j \gtrsim 10^{1.5} r_s$. Note that the
characteristic jet scale $r_j$ and the jet (non-thermal) luminosity $L_{NT}$ are
considered as independent variables, measured in units of $r_s$ and
$L_{E}$. 

The break seen at few keV is attributed to emission from electrons at
$\gamma_m$, either via synchrotron emission or Comptonization of the
disk photons. As these two scenarios lead to different constraints on
the model parameters, we treat each one separately. 

In the calculations below, we scale the values of the model parameters
to the reference values relevant for the 2000 outburst of XTE
J1118+480. This is done for ease of comparison with the numerical
results presented in \S\ref{numerics}. As the discussion is
general and the dependence on the spectral breaks and luminosities are
given, generalization to any source is readily obtained.

\subsection{Synchrotron-dominated model}
\label{sec:synchrotron}

Electrons with Lorentz factor $\gamma_m$ radiate synchrotron photons
at characteristic observed energy $\epsilon_m^{ob} = (3/2) \hbar (q B / m_e c)
\gamma_m^2 \D= 1.75 \times 10^{-8} B \gamma_m^2 \D \;\;\eV$. Here,
$q$ is the electron's charge, and $\D \equiv [\gamma_j (1- \beta_j \cos
\theta)]^{-1} \gtrsim 1$ is the Doppler shift ($\theta$ is the angle
to the line of sight). The requirement that $\varepsilon_{\min} \simeq
2 \keV$ thus constraint the values of $B
\gamma_{\min}^2$\footnote{For mildly relativistic jets, $\gamma_j
  \gtrsim 1$, the Doppler shift is not strongly constrained.}. A
further constraint on $B$ is added by the physical requirement that
$\epsilon_e \leq 1$, which provides an upper limit on the value of
$\gamma_m$ via equation \ref{eq:gamma_m}. 

The second requirement is that the cooling break, $\nu_c$, which is
the characteristic frequency of synchrotron emission from electrons
having Lorentz factor $\gamma_c$ is obscured by the disk photons,
hence $\varepsilon_c^{ob} = \hbar \nu_c^{ob} \lesssim 0.3
\keV$. Finally, we require that synchrotron emission is the dominant
cooling mechanism, which translates into $u_B \gg u_{th}$. Combined
with the requirement on the optical depth (eq. \ref{eq:tau}), one
obtains the set of constraints
\beq
\ba{llcll}
1.a. &  \varepsilon_m^{ob} \simeq 2 \keV :
\quad B \gamma_m^2  \simeq  1.2
\times 10^{11} \D^{-1} {\rm \, G} 
\nonumber \\
1.b.&  \epsilon_e \leq 1:
\quad \quad \quad \;  B  \gtrsim  3.4 \times 10^4 \D^{-1} {\rm
  \, G} \nonumber \\
2. & \varepsilon_c^{ob} \lesssim 0.3 \keV:\nonumber \\
& \; B  \gtrsim 4.3 \times 10^4 
(\beta_j/\gamma_j)^{2/3}  (r_j/10r_s)^{-2/3} \D^{1/3} {\rm \, G}
\nonumber \\
3.& u_B \gg u_{th}: \nonumber \\
& \; B  \gg  1.3 \times 10^6 \beta_j^{-1/2}
\gamma_j^{-1} \left({L_{th} / 3.2\times 10^{-3} L_E}\right)^{1/2}
\times \nonumber \\
& \quad \quad  \; \left({r_j / 10  r_s}\right)^{-1} {\rm \, G} \nonumber \\
4. & \tau_{\gamma,e^\pm} \lesssim 1: 
 \quad \quad  (r_j/10 r_s)  \gtrsim  1
\beta_j^{-3/2} \gamma_j^{-1/2}
\ea
\label{eq:constraints1}
\eeq
In equation \ref{eq:constraints1}, the thermal (disk) luminosity is normalized
to $3.2 \times 10^{-3} L_E$, as is in the 2000 outburst of XTE
J1118+480; see further discussion in \S\ref{numerics} below.

One thus obtains strong constraints on the value of $B \gtrsim 10^6$~G
and the characteristic scale $r_j \gtrsim 10^{1.5} r_s$. While the
value of $\epsilon_e$ is not strictly limited from below, a high value
of $\epsilon_e \lesssim 1$ is generally preferred, since lower values
of $\epsilon_e$ implies, via eq. \ref{eq:constraints1} (1.a.)
stronger magnetic field, which may be challenging to produce.  High
value of $\epsilon_e$ implies, in turn, high radiative efficiency -
namely, that the jet kinetic luminosity is similar to the non-thermal
luminosity seen.  We further note that there is no constraint on the
value of $\xi_{pl}$ set by the data below $\epsilon_m^{ob}$.

\subsection{Compton-dominated model}
\label{sec:Compton}

Alternatively, the break seen at $\sim {\rm few} \keV$ may be
attributed to inverse Comptonization of the disk photons by the jet
electrons. As the angle-averaged Comptonized energy is $\varepsilon_m
= (4/3) \gamma_m^2 \varepsilon_{in}$ where $\varepsilon_{in}^{ob}
\simeq 0.1 \keV$ is the energy of the disk photons (using the value
from the 2000 outburst of XTE J1118+480), in this scenario the
characteristic electrons Lorentz factor is $\gamma_m \lesssim
10$. This result, in turn, implies $\epsilon_e \gtrsim 3\times 10^{-3}$, and
$L_k \lesssim L_E$.

In this scenario, the cooling is dominated by Comptonization of the
disk photons, $u_{th} \gg u_B$. Finally, one requires the cooling time
to be shorter than the dynamical time, which implies $\gamma_c \gtrsim
1$. Combined together, these conditions give
\beq
\ba{llccl}
1. &  \varepsilon_m^{ob} \simeq 2 \keV: \quad \epsilon_e  \gtrsim  3
\times 10^{-3}
\nonumber \\
2. & \gamma_c \simeq 1:  \quad \quad \quad  \,  (r/10 r_s)  \sim  2 \beta_j^{-2}
\gamma_j^{-1} \left({L_{th} / 3.2\times 10^{-3} L_E}\right) 
\nonumber \\
3.& u_B \ll u_{th}: \nonumber \\
& \; B  \ll  1.3 \times 10^6 \beta_j^{-1/2}
\gamma_j^{-1} \left({L_{th} / 3.2\times 10^{-3} L_E}\right)^{1/2}
\times \nonumber \\
& \quad \quad \;  \left({r_j / 10
    r_s}\right)^{-1} {\rm \, G} \nonumber \\
4. & \tau_{\gamma,e^\pm} \lesssim 1: \quad \quad  (r_j/10 r_s)  \gtrsim  1
\beta_j^{-3/2} \gamma_j^{-1/2}.
\ea
\label{eq:constraints2}
\eeq

We stress again that while the derived constraints on the parameters
values are obtained using the data from the 2000 outburst of XTE
J1118+480, the numerical values for other sources are easily obtained
by inserting the values of the break energies and the thermal and
non-thermal luminosities in equations \ref{eq:constraints1},
\ref{eq:constraints2}.

\subsection{Numerical results}
\label{numerics}

\begin{figure}
\plotone{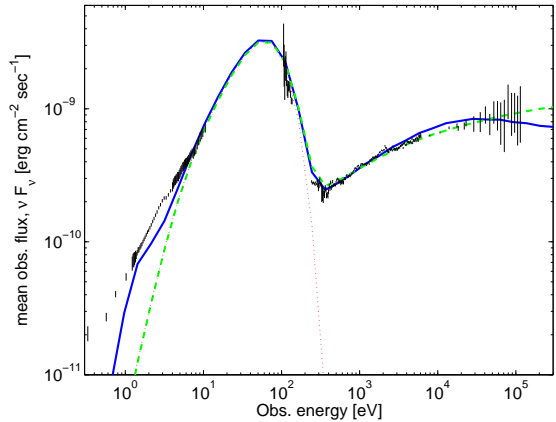}
\caption{Example of the numerical results, on top of the data of the
  2000 outburst of XTE J1118+480. The solid (blue) curve represent the
  synchrotron-dominated model, while the dashed (green) curve
  represents the inverse Compton-dominated model. A power law-
  dominated distribution of accelerated electrons ($\xi_{pl} = 1$) is
  assumed.  The values of the other model parameters are given in the
  text. No attempt was made to statistically fit the data, but rather
  this plot demonstrates the ability of this transient jet model to
  reproduce the basic features of the observed X-ray spectra. The
  dotted (red) line is (multi-color) black body component, originating
  from the accretion disk. These photons (together with the
  synchrotron photons) serve as seed photons for Compton scattering by
  the energetic electrons.}
\label{fig:1}
\end{figure}

\begin{figure}
\plotone{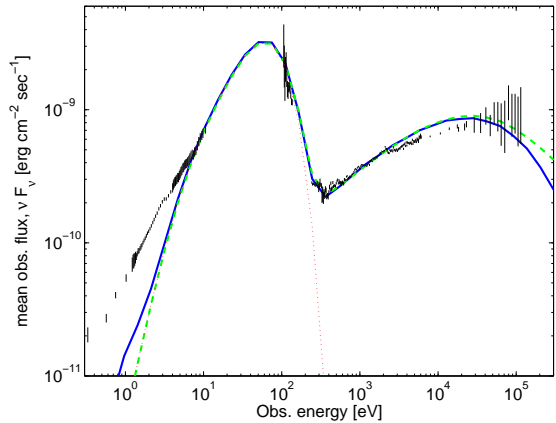}
\caption{Similar to Figure \ref{fig:1}, but with only a small fraction
  ($\xi_{pl} = 0.01$) of particles accelerated into a power law, the
  rest maintain a Maxwellian distribution. As explained in the test,
  reasonable fits are obtained in this scenario as well, due to the
  fact that in the fast cooling regime, a universal spectrum $F_{\nu}
  \propto \nu^{1/2}$ is produced regardless of the initial electrons
  spectra, and that at high energies, the finite width of the
  Maxwellian enables fitting the data. In this scenario, the fit to
  the data does not uniquely determine the value of power-law index
  $p$.  }
\label{fig:2}
\end{figure}

The analytical constraints derived in equations \ref{eq:constraints1},
\ref{eq:constraints2} provide good approximations for the required
conditions from jet model needed in fitting an observed spectrum.
However, due to the complexity of the problem, in particular the
possible creation and annihilation of a large number of pairs, a
numerical approach is needed in order to validate the analytical
approximations and provide more accurate constraints.  Here, we use a
time-dependent model, based on the code developed by
\citet{PW05}. This code solves self-consistently the kinetic equations
that govern the time-dependent energy distribution and radiation from
particles inside the jets, following all relevant radiative
processes. These include synchrotron emission, Compton scattering,
pair production and annihilation that follow particle acceleration at
the base of the jet. An example of the numerical results are presented
in Figures \ref{fig:1}, \ref{fig:2} on top of the data from the 2000
outburst of XTE J1118+480. We stress that no attempt was made to
obtain the statistical best fit to the data, but only to demonstrate
the possibility of reproducing the observed X-ray spectral slopes in
the two model scenarios considered.

In Figures \ref{fig:1} and \ref{fig:2} we show the results of our
exploration of two scenarios, strong particle acceleration ($\xi_{pl}
=1$) and weak particle acceleration ($\xi_{pl} =0.01$),
respectively. In both figures, the solid (blue) line represents the
numerical result for the synchrotron-dominated model. We chose
representative parameters consistent with a mildly-relativistic jet
($\beta_j = 0.4$), similar to the value chosen in \citet{FB95,
  MFF01}. The disk is fitted with a multi-color black body component,
with $L_{th} = 0.0032 L_E$, and inner disk temperature $T_{in} = 5
\times 10^{-11}$~erg, giving an inner disk radius of $r_{in}/r_s =
326$.  We chose parameters values consistent with the constraints set
in equation \ref{eq:constraints1}, such as magnetic field $B=3\times
10^6$~G and $r_j/r_s = 200$. The value of $\epsilon_e$ is close to
equipartition: $\epsilon_e = 1.0$ in Figure \ref{fig:1} and
$\epsilon_e = 0.5$ in Figure \ref{fig:2}. The resulting spectrum is
very similar to the X-ray spectrum of XTE J1118+480, for $L_k = 0.0045
L_E$ (Figure \ref{fig:1}) and $L_k = 0.005 L_E$ (Figure \ref{fig:2}) .
In the results presented in both figures, the power law chosen above
$\gamma_m$ is $p=2.0$, consistent with Fermi acceleration models. We
note that in the $\xi_{pl} = 0.01$ scenario presented in Figure
\ref{fig:2}, the high energy part of the spectrum is fitted with
particles at the peak of the Maxwellian. Reasonable fits are obtained
due to the finite width of the Maxwellian distribution. In this
scenario, however, the exact value of $p$ cannot be determined from
the data.

For these parameter values, the flow is (by definition) magnetic
dominated: the energy density in the magnetic field, $u_B \equiv B^2/8
\pi \simeq 3 \times 10^{11} \; {\rm \erg \, \cm^{-3}}$ is much larger
than the kinetic energy density, $u_k = L_k / \pi r_j^2 \beta_j c
\gamma_j^2 \simeq 5 \times 10^8 \; {\rm \erg \, \cm^{-3}}$.

The result of a Compton-dominated model is presented by the dashed
(green) line in Figures \ref{fig:1} and \ref{fig:2}. The values of the
disk parameters and the jet velocity are similar to the ones used in
the synchrotron model.  The values of the free parameters of the jet
model chosen, $r_j/r_s = 25$ and $\epsilon_e = 0.029$ (Figure
\ref{fig:1}), $\epsilon_e = 0.01$ (Figure \ref{fig:2}) reproduce the
data well for $L_k = 0.15 L_E$ (Figure \ref{fig:1}), $L_k = 0.2$
(Figure \ref{fig:2}) , and weak magnetic field. Here we chose $B =
10^3$~G, which implies matter dominated outflow - $u_B/u_{k} \simeq 3
\times 10^{-8}$. As explained above, the exact value of the magnetic
field is unimportant as long as the constraints in equation
\ref{eq:constraints2} are fulfilled. In this scenario as well, for
$\xi_{pl} = 1$ a power law index $p=2.0$ results in very good fit to
the data.

\section{Summary and discussion}
\label{sec:summary}

In this paper, we present a new jet-dominated model that is able to
reproduce the main spectral properties seen at the X-rays of many XRBs
in the hard state. Our key motivation is the spectral break which is
often seen at $\sim$~few \keV - few tens \keV, and (within these
sources) the nearly universal spectral slope, $F_\nu \propto \nu
^{1/2}$ observed below this break. This spectral slope is a natural
outcome of emission from electrons whose energy distribution is
determined by rapid radiative cooling, following acceleration at the
jet base. The rapid cooling can result from either synchrotron
emission or Comptonization of the disk photons, or as likely, a
contribution from both processes.  We derive in equation
\ref{eq:constraints1}, \ref{eq:constraints2} the required constraints
on the free model parameters for both these scenarios, and demonstrate
the resulting spectrum in Figures \ref{fig:1} and \ref{fig:2}.

The key difference between our model and earlier jet models is that
here we self-consistently consider the temporal variation of the
particle distribution due to the radiative cooling, and consider those
variations in the spectral calculations. Thus, we are able to make a
clear separation between the acceleration process and the cooling
processes; namely, we do not assume that the energy lost by the
electrons is necessarily fully replenished by any heating source.
Indeed, in recent years it has become clear that this decoupling
between acceleration and cooling may hold the key to understanding the
spectral properties of XRBs. As a result, in recent years several
numerical models which consider this separation have been constructed
in the study of XRBs spectra \citep[e.g.,][]{BM08, MMF09}. In essence,
these models are very similar to the numerical model used here. A key
advantage of our model is its ability of solving the kinetic equations
over many orders of magnitude in time and energy scales, which makes
it ideal for producing broadband spectra.

As indicated in Figures \ref{fig:1} and \ref{fig:2}, both synchrotron-
and Compton-dominated scenarios can in principle reproduce the X-ray
spectrum of a typical source showing the predicted break. The values
of the free model parameters are, however, significantly different in
the two scenarios. While the Compton dominated scenario requires
$r/r_s \lesssim 10^{1.5}$, $L_k \simeq L_E$ and a relatively weak
magnetic field, the synchrotron dominated scenario better fits the
data with $r/r_s \gtrsim 10^2$, $L_k \lesssim 10^{-2} L_E$ and a
strong magnetic field, $B\gtrsim 10^6$~G.  In our opinion, at least
for XTE~J1118+480, the data favor a synchrotron-dominated scenario,
similar to the conclusions of \cite{MFF01,Maitra+09}. First, the
characteristic scaling $r_j$ is similar to the inner (truncated) disk
radius, $r_{in}$. Thus, obviously, this radius marks a physical
transition in the properties of the flow, which can result from
development and launching of jets. Second, the non-thermal radio
spectra observed in the hard states provide clear indication for
jet-synchrotron emission. Emission at the radio band is expected once
the plasma reaches a scale much larger than $r_j$ (and with
corresponding much weaker magnetic field). As the magnetic field is
expected to decay along the jet (due, e.g., to Poynting-flux
conservation), unless there is a generation of strong magnetic fields
at some scale $r\gg r_j$, a strong magnetic field at the jet base
decays along the jet, and thus the same field may be the source of the
radio emission, further along the jet. While studying the full
connection between the X-ray spectrum and the radio properties is left
for future work, we note that such a connection may exist, and may
lead to very interesting consequences \citep{PC09,CP09}.

The numerical results presented in Figures \ref{fig:1} and \ref{fig:2}
show that the change in the spectral index (the spectral break) is
gradual, extending over about an order of magnitude in energy.  This
result originates from the broad band nature of the emission
processes. Thus, in particular for scenarios in which the
characteristic breaks $\varepsilon_m^{ob}$ and $\varepsilon_c^{ob}$
are close to each other in energy, a gradual change of the slope is
expected, providing a natural explanation for the somewhat softer
slope observed below the break in several objects.

Similarly, a general trend seen in many objects is a softening of the
spectrum with increase of the luminosity, marking the transition to
the soft state \citep{Migliari07, Dunn+11}. This result can be
understood in the framework of the model presented here as follows. As
the luminosity increases, the conditions at the acceleration region
are likely to vary. The strength of the magnetic field $B$ as well as
the acceleration efficiency ($\epsilon_e$) are likely decreasing, as
is indicated by the lack of radio emission in the soft state. In both
the synchrotron and the Compton dominated models, the break energy
$\varepsilon_m^{ob}$ depends on $\epsilon_e^2$, and in the magnetized
model also on $B$. Thus, as a result of the changing conditions,
$\varepsilon_m^{ob}$ is shifted to lower energies, and the break is
eventually obscured by the disk photons.  At this stage, the X-ray
spectrum represents the emission above the spectral break, which is
softer.

Another interesting result of this work is the ability to reproduce
the observed spectra with acceleration processes that produce a power
law index $p\approx2$ for the accelerated electrons, under the
assumption that $\xi_{pl} \simeq 1$. While a full theory of the
acceleration process still does not exist, a plethora of observational
evidence exists that indicates a canonical power law index of the
accelerated electrons. These include evidence from AGN jets,
supernovae remnants and GRB shock waves, all showing canonical index
$p\gtrsim 2.0$. Our results are thus consistent with these findings,
and do not require a modification of this law. While the quality of
the data above the spectral break in many objects is not very good, we
note that often a roughly flat power flux ($\nu F_\nu \propto \nu^0$)
is seen, further strengthening this result.

While in recent years significant progress in understanding particle
acceleration in shock waves has been made, there is currently a
discrepancy between the theoretical expectation of $\xi_{pl} \ll 1$
\citep[e.g.,][]{AB05, Spit08, Caprioli+10, SS11} and the
phenomenological fittings indicating higher value of $\xi_{pl} \simeq
1$ \citep[e.g.,][]{WG99, FWK00, FW01, PK02, MNW05}. This discrepancy
can perhaps be explained by the fact that the theory is not yet
complete.  Since this topic is outside the scope of this work, we have
explored the consequences for both scenarios.

Specifically, we have demonstrated that in the alternate scenario in
which most accelerated electrons maintain a Maxwellian distribution
($\xi_{pl} \ll 1$), good fits to the data can still be obtained.  The
high energy part of the spectrum is explained by emission from
particles at the peak of the Maxwellian, although this scenario does
not constrain the value of the power-law index $p$.

While we show here that a jet model is able to reproduce the X-ray
spectra, we note that possible contribution to the X-ray emission may
arise from the inner parts of the disk (RIAF models).  Observationally
there are some new results that hint at an interplay of the two
processes as a function of the total luminosity \citep{YC05,
  Russell+10, Plotkin+11, Gandhi+11}, where RIAFs may begin to
dominate at the higher hard state luminosities.

From a theoretical point of view, the conditions in the inner part of
the inflow may be similar to the conditions in the inner part of the
outflow \citep[see discussion in ][]{MNW05}. Hence, a clear separation
between the inner disk contribution and the inner jet contribution may
be difficult.  Several methods may be used to separate the disk and jet
contributions. First, a clear connection between the X-ray emission
and the radio/NIR emission, which must have a jet origin, could provide
evidence for jet dominated model, provided that the physical
conditions at both emission sites (e.g., the magnetic field or the
number of radiating particles) are matched.   Such a connection
already seems to be indicated by the observational papers cited above.

A second method is based on the requirement that the optical depth to
scattering of the disk photons needs to be not much larger than unity,
otherwise the spectral shape of the disk will be altered. Thus, if the
disk accretion time is significantly larger than $\sim r_j/c$, this
could enable multiple Compton scatterings between the electrons and
the disk photons, which may introduce a different characteristic from
inflow emission. A full treatment of this idea is left for a future
work.  Finally, the best way to discriminate between disk and jet
models may be using X-ray polarization measurements. Both synchrotron
emission and non-saturated Compton scattering are highly polarized,
hence jet emission is expected to be polarized. On the other hand, in
RIAF models the polarization signal is expected to be weaker, due to
averaging of the synchrotron signal from different parts of the disk,
as well as contribution from free-free emission from the outer
accretion flow.  We thus expect that proposed future X-ray
polarization missions, such as the POLARIX mission \citep{Costa+10} or
the {\it Gravity and Extreme Magnetism Small Explorer} (GEMS) mission
\citep{Jahoda10} may play a significant role in determining the
relative contributions from the disks and the jets in producing the
X-ray spectra.

\acknowledgments
We would like to thank Dipankar Maitra, Shmulik Balberg, Ramesh
Narayan, J\"orn Wilms and Piergiorgio Casella for useful discussions
and comments. A special thanks for Dipankar Maitra for providing us
the reduced data from the 2000 outburst of XTE J1118+480 (from Hynes
et al. 2000).  SM is grateful for support from a Netherlands
Organization for Scientific Research (NWO) Vidi Fellowship, and the
European Community's Seventh Framework Programme (FP7/2007-2013) under
grant agreement number ITN 215212, ``Black Hole Universe''.

\end{document}